\documentclass[sigconf]{acmart}
\usepackage{makecell}
\usepackage{multirow}
\AtBeginDocument{%
  }

\copyrightyear{2025}
\acmYear{2025}
\setcopyright{rightsretained}
\acmConference[CHI EA '25]{Extended Abstracts of the CHI Conference on Human Factors in Computing Systems}{April 26-May 1, 2025}{Yokohama, Japan}
\acmBooktitle{Extended Abstracts of the CHI Conference on Human Factors in Computing Systems (CHI EA '25), April 26-May 1, 2025, Yokohama, Japan}\acmDOI{10.1145/3706599.3719900}
\acmISBN{979-8-4007-1395-8/2025/04}

\begin{document}

\title{Negotiating the Shared Agency between Humans \& AI in the Recommender System}

\author{Mengke Wu}
\email{mengkew2@illinois.edu}
\affiliation{
    \institution{University of Illinois Urbana-Champaign}
    \department{School of Information Sciences}
    \city{Champaign}
    \state{Illinois}
    \country{USA}
}

\author{Weizi Liu}
\email{weizi.liu@tcu.edu}
\affiliation{
    \institution{Texas Christian University}
    \department{Bob Schieffer College of Communication}
    \city{Fort Worth}
    \state{Texas}
    \country{USA}
}

\author{Yanyun Wang}
\email{mia.wang@colorado.edu}
\affiliation{
    \institution{University of Colorado Boulder}
    \department{College of Media}
    \city{Boulder}
    \state{Colorado}
    \country{USA}
}

\author{Mike Yao}
\email{mzyao@illinois.edu}
\affiliation{
    \institution{University of Illinois Urbana-Champaign}
    \department{Institute of Communications Research}
    \city{Champaign}
    \state{Illinois}
    \country{USA}
}

\renewcommand{\shortauthors}{Wu et al.}

\begin{abstract}
Smart recommendation algorithms have revolutionized content delivery and improved efficiency across various domains. However, concerns about user agency arise from the algorithms’ inherent opacity (information asymmetry) and one-way output (power asymmetry). This study introduces a dual-control mechanism aimed at enhancing user agency, empowering users to manage both data collection and, novelly, the degree of algorithmically tailored content they receive. In a between-subject experiment with 161 participants, we evaluated the impact of varying levels of transparency and control on user experience. Results show that transparency alone is insufficient to foster a sense of agency, and may even exacerbate disempowerment compared to displaying outcomes directly. Conversely, combining transparency with user controls—particularly those allowing direct influence on outcomes—significantly enhances user agency. This research provides a proof-of-concept for a novel approach and lays the groundwork for designing more user-centered recommender systems that emphasize user autonomy and fairness in AI-driven content delivery.
\end{abstract}




\maketitle

\section{INTRODUCTION}
Behavioral targeting and predictive algorithms have revolutionized information dissemination by enhancing efficiency and fundamentally altering the landscape of media content production and distribution. These algorithms leverage user data to predict their interests, then automatically select and deliver relevant content \cite{pandey2011learning, yoneda2019algorithms}. This process finds extensive use across domains, encompassing social media, news platforms, forums, streaming services, e-commerce, and more. While much research has investigated the construction and impact of these recommender systems (RS), they have also raised discussions and challenges concerning human-AI interactions and the pursuit of human-centered design \cite{ehsan2021operationalizing, shin2019role}. A common theme is the opacity of algorithmic decision-making, which operates as an inscrutable “black box” that users can only access and perceive the provided recommendations \cite{pasquale2015black, burrell2016machine}, lacking explanations and involvement in this targeting and personalization process \cite{felzmann2020towards, weller2019transparency}. This leads to two key issues: \textbf{1) information asymmetry}, which diminishes users’ perceptual agency by providing limited understandings of the reasoning behind recommendations \cite{kizilcec2016much, rader2018explanations, sonboli2021fairness}, and \textbf{2) power asymmetry}, which erodes users’ behavioral agency by restricting their control over algorithmic decisions \cite{cheng2023overcoming, molina2022ai, sundar2020rise}. These asymmetries threaten user-centered design principles and can lead to manipulation, information blockage, or loss of trust \cite{conover2011political, kizilcec2016much, nguyen2014exploring, ruhr2023intelligent}.

Recent advancements, particularly in AI transparency and explainable AI (XAI) \cite{kizilcec2016much, rader2018explanations}, have sought to reduce information asymmetry by making data usage more transparent and controllable (e.g., cookie disclaimers \cite{segijn2021literature}). Practices in human-AI collaborative decision-making (HACD) have also aimed to distribute power to users by granting controls over permissions \cite{burrell2016machine, segijn2021literature} and providing avenues for algorithm improvements \cite{lai2023towards, molina2022ai}. However, these efforts often fall short in addressing power asymmetry and users remain limited in controlling algorithmic targeting outcomes. This underlines the need for more comprehensive approaches that integrate both transparency and control, ensuring users not only understand how decisions are made but also have the ability to influence them. This study intends to advance HACD by proposing a dual-control mechanism that enhances users' sense of agency regarding personal data collection and usage while also allowing users to adjust the degree of personalized content they wish to receive. Meanwhile, we explore how varying levels and types of control in media content personalization impact user evaluations of the system. Through a between-subject experiment embedding different recommendation prototypes, we empirically investigated the impact of this mechanism. Participants interacted with the assigned prototype and completed a post-assessment. The findings contribute to the broader discourse on algorithmic fairness, user autonomy, and human-AI collaboration. We also provide practical guidelines for refining RS that balance algorithmic efficiency with user agency, ultimately empowering users with greater control, understanding, and trust in the digital age.

\section{RELATED WORK} \label{lit}

\subsection{Information Asymmetry} 
Information asymmetry exists in many algorithm-based RS, pertaining to the discrepancy between what output users could see (the recommendations) and what they couldn’t (how recommendations were generated). Such asymmetry often prevents users from fully comprehending the underlying processes and leads to a lack of perceptual user agency, which is misaligned with the pursuit of “interpretability” and “explainability” in complex algorithms. Scholars have emphasized the significance of making algorithmic RS transparent and interpretable, especially from the user perspective \cite{cramer2008effects, kizilcec2016much, sonboli2021fairness}. This is further linked to the situation awareness for autonomous systems and AI to ensure effective interaction and oversight \cite{endsley2023supporting}. To tackle this issue, multiple strategies have been proposed, such as showing 1) source transparency, which clarifies why and where such recommendations come from \cite{felzmann2020towards}; and/or 2) process transparency, with a focus on Explainable AI (XAI) that elucidates the system’s behavior \cite{rader2018explanations, ras2018explanation}. Cookie disclaimers are the common ways to enhance information transparency, which inform users about the data collection and its role in generating personalized content \cite{millett2001cookies, segijn2021literature}.

Guidelines have also been put forth to guide the development and design of algorithm-based systems to greater transparency and explainability. Liao et al. (2020) proposed a question bank \cite{liao2020questioning} and Microsoft developed the HAX design guidelines \cite{amershi2019guidelines}, both aimed at prompting considerations on system transparency when creating user-centered XAI. Usability guidelines for user interface and XAI system design have also been advocated to address users’ needs for learning “what to explain” and “how to explain” \cite{eiband2018bringing, wolf2019explainability}.

\subsection{Power Asymmetry}
While information asymmetry
might be perceptual in nature, power asymmetry affects their agency behaviorally.
Power asymmetry in AI-driven RS underscores an imbalance between user control and algorithmic authority. Sundar (2020) recognized the inherent tension between human and machine agency, identifying the loss of agency as a major factor driving fears about automation and advocating for the human-AI synergy concept, stressing the need for users' direct role in shaping algorithms to meet their needs \cite{sundar2020rise}.
In contrast, persistent power asymmetry can hinder user adoption, as lower perceived control reduces perceived usefulness and the intention to engage with the system \cite{chen2018app}. Proactive and reactive relationships between users and algorithms have also been discussed, particularly in RS. Proactive interactions involve machine agency predicting user needs and presenting content based on analyzed patterns, while reactive interactions exhibit user agency that responds to explicit user requests by personalized suggestions \cite{zhang2019proactive}.

\subsubsection{Existing User Control in RS}
User control mechanisms are a vital response to power asymmetry, essential for enhancing transparency, trust, personalization, and user satisfaction. They also address ethical concerns, such as accountability and democratic participation \cite{harambam2019designing}. Emerging concepts like interactive transparency \cite{molina2022ai} and human-AI collaborative decision-making (HACD) \cite{lai2023towards, schemmer2022meta} emphasize respecting user feelings and involving users in
algorithm's decision-making process,
thereby bolstering their trust and perceived control toward the system.

\begin{figure*}
    \centering
    \includegraphics[width=1\linewidth]{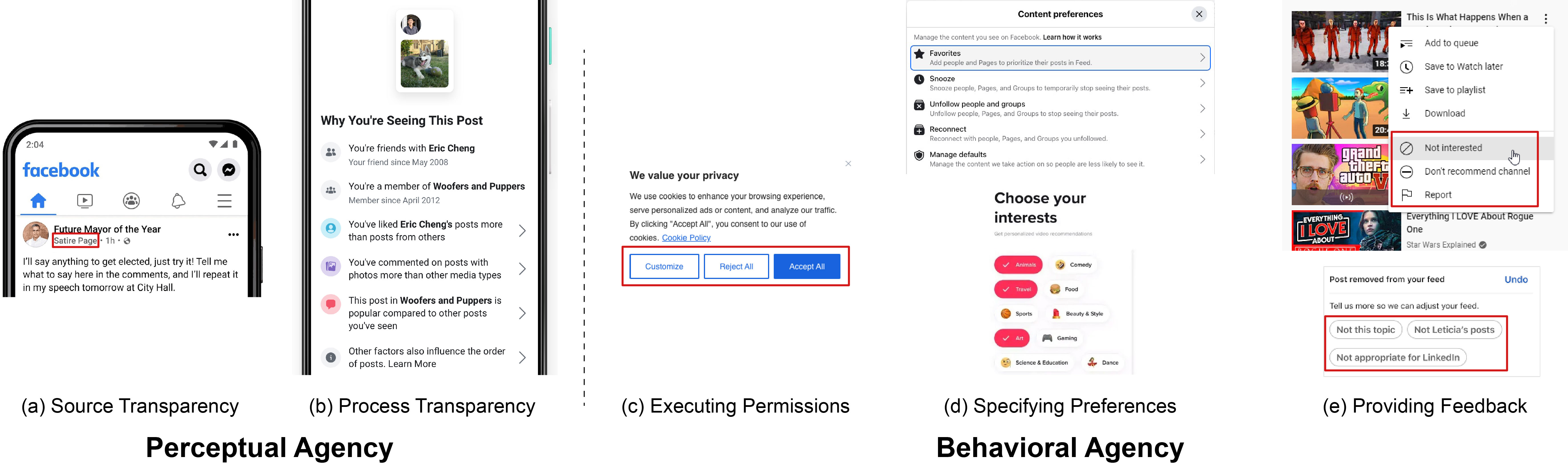}
    \caption{Examples of Different Types of Agency. Perceptual Agency (Left): (a) Source Transparency, (b) Process Transparency; Behavioral Agency (Right): (c) Executing Permissions, (d) Specifying Preferences, (e) Providing Feedback.}
    \label{fig:current}
\end{figure*}

Behavioral agency centers around users’ collaborations on 1) granting or withholding permissions (e.g., decline cookies) \cite{berens2024cookie, segijn2021literature}; 2) specifying preferences for personalized recommendations \cite{rashid2002getting, zheng2022perd}; and 3) providing feedback to refine decision-making processes \cite{cheng2023overcoming, molina2022ai}. For example, studies have introduced features for users to understand and adjust how their data influences recommendations by inspecting and modifying their profiles, thereby improving system transparency and trust \cite{bakalov2013approach}. This has been further expanded by designing interfaces for preference specification, recommendation adjustments, and feedback, creating more interactive and responsive experiences \cite{jannach2017user, jannach2019explanations}. Other efforts include adjustable parameters (e.g., popularity, recency), which offer immediate influence over recommendations and enhance satisfaction \cite{harper2015putting}. Meta-recommender systems also empower users by combining multiple recommendation sources, providing greater personalization and control over the recommendation process \cite{schafer2002meta}. Figure \ref{fig:current} presents examples of different types of user agency in current practices.

Studies also explore how users perceive control and the psychological factors behind it. Value-added functions, like adapting control interfaces to individual traits, increase recommendation acceptance \cite{jin2020effects}. Combining control with visual explanations further enhances transparency and user satisfaction \cite{tsai2021effects}. In conversational RS, adaptability, understanding, and responsiveness are critical for fostering a sense of control \cite{jin2021key}. Well-designed control mechanisms can also reduce cognitive load and improve engagement, directly strengthening users’ perceived control \cite{laban2020effect}.

\subsubsection{Opportunities for Shaping Power Dynamics}
Despite these developments, challenges and concerns like filter bubbles, over-specialization, and accuracy-drivenness remain prevalent in AI RS, leading to predictable and monotonous outputs with limited diversity, negatively affecting user experience \cite{rowland2011filter, ziarani2021serendipity, adamopoulos2014over}. Current user controls largely focus on filtering and refining algorithmic results, as mentioned above. While efforts are made to optimize backend algorithms for diversity, novelty, and serendipity in recommendations \cite{kaminskas2016diversity, ziarani2021serendipity}—few consider externally from the user side in asking how much AI-recommended content they need before intervening in algorithmic decisions. Questions also remain about whether systems should proactively solicit user input regarding this perspective and how such interactions should be designed to optimize user control and experience. This study aims to advance the theoretical construct of behavioral agency by applying it to design applications. We propose an intuitive mechanism allowing users to actively regulate the degree of algorithmically recommended content. Through this approach, we ask:

\textbf{RQ1:} Does the ability to control the degree of AI-recommended content enhance user agency in the recommender system?

Moreover, research mainly focused on individual aspects of agency. For instance, some researchers examined how perceptual agency influences decisions about action outcomes \cite{desantis2016agency}, while others emphasized the role of behavioral agency in shaping user experiences \cite{coyle2012did}. However, few studies have holistically examined the comparative effects of perceptual and behavioral agency. This study aims to bridge this gap by integrating both types to investigate their interplay and impacts on user experiences. We seek to understand:

\textbf{RQ2:} How do user perceptions, experiences, and attitudes vary if given different types of agency over recommendation algorithms?

\section{METHODS}
To enhance user engagement in the research process, this study selected a web-based news platform as the sample RS for its strong alignment with the societal focus and information dissemination goals of the study. A pretest survey was conducted to identify suitable news snippets as recommendation stimuli. The main study employed a between-subject experiment with varied flow prototypes to examine participants’ perceptions, experiences, and attitudes toward this RS prototype. As the first empirical study in a planned sequence, the findings presented here serve as a foundation for future research, which will involve a more comprehensive and in-depth investigation.

\subsection{Pretest}
To ensure politically neutral yet engaging recommendation results, a pretest was conducted to select news snippets. We aimed for content with no obvious tendency and would be equally relevant and interesting to all participants to mitigate content-driven bias. Using ChatGPT (4.0), we generate 50 news snippets with the prompt: \textit{“help me generate 50 news titles that do not have any obvious political leaning but can be interpreted from a partisan perspective, add one sentence summary after each title.”} Artificially generated snippets allowed better control over potential biases from real-world headlines that may reflect the editorial or political leanings of their sources. This approach also helped to focus the study on user engagement with the content itself, rather than any prior familiarity or perceptions linked to news source. We manually reviewed the snippets to ensure they were balanced, appropriate, and free from uncommon content that might skew interpretation. The snippets were then paired randomly by the survey system. In each instance, the system randomly selected two snippets from the remaining pool, excluding those already chosen, and presented them for participants to choose the one they found more interesting, resulting in 25 pairwise choices. Demographic questions, especially their political stance, were also collected.

55 participants from the Amazon Mechanical Turk platform were recruited for the pretest (61.8\% males and 38.2\% females; average age of 31.6 with \textit{SD} = 7.51). Data analysis for the pre-test followed four steps: 1) calculating average choice frequency by political stance (Democrat/Republican); 2) determining the median of these averages; 3) identifying snippets with minimal frequency differences as those with no clear tendency to parties (\begin{math} |Democrat-Republican|\leq1 \end{math}); and 4) within the identified ones, selecting those with above-median average frequency as of higher interest to all parties. This process yielded 12 balanced, highly-interest snippets as our recommendation results for all flow prototypes.

\subsection{Main Study}

\subsubsection{Study Sample \& Procedure}
Participants were recruited from a large U.S. university using an online research participation pool, with all participants receiving extra research credits. A total of 161 participants (24.8\% males, 75.6\% females) provided valid responses. The average age of the sample was \textit{M} = 20.80 (\textit{SD} = 3.10).

After providing informed consent, participants began the study by completing demographic questions, including their political stance. This was designed to create the illusion of personalized news selection when they later encountered the pre-selected news snippets with general relevance and interest. By collecting political information upfront, we intended to simulate an environment where the news recommendations appeared tailored to participants' inputs, thus enhancing the perceived relevance and realism of the system. Participants were then randomly assigned to one of five conceptualized prototypes of news recommendation flows, each representing a different type and level of agency over the recommendation process. The final recommended news snippets remained the same (see Section \ref{flow}). The flow interaction for each participant was one-way and non-recurring. After experiencing the flow and viewing the recommendations, participants returned to the survey to complete a post-interaction assessment of their experience with that recommendation flow.

\subsubsection{Algorithm Outcome Control}
Within the dual control mechanism, we introduced the algorithm outcome control (AOC) slider to novelly give users control over the degree of algorithmic recommendations they receive based on their data (0\% means opting out of tailored content and not seeking personalized recommendations). By allowing users to shift their preferences between precision (tailored) and exploration (non-tailored), the AOC slider seeks to rebalance user power and foster a more user-centered and collaborative recommendation system. This tool aligns with the growing focus on human-AI synergy \cite{sundar2020rise} and interactive transparency \cite{molina2022ai}, providing a tangible mechanism for users to influence the algorithmic process and their experiences.

\subsubsection{Flow Design} \label{flow}
The flow design integrates elements based on the two types of agency discussed in Section \ref{lit}. To address information asymmetry and increase perceptual agency, we operationalized algorithm transparency (T) by providing a cookie disclaimer disclosing the use of user data and specifying the types of data collected to improve recommendations. Regarding power asymmetry, we introduced two levels of control to enhance behavioral agency: user data control (UDC) and our to-be-tested idea AOC. Following established practices, UDC afforded users the choice of accepting or declining the cookies (declining meant refusal to data collection).

We conceptualized five flow conditions, ranging from no control to full control. Flow 1 (None) served as a baseline, displaying recommended results without any visible mechanism cues. Flow 2 (Only T) presented transparency with a view-only cookie disclaimer, providing perceptual agency. Building on Flow 2, Flow 3 (T + UDC) enabled participants to accept or decline data collection, while Flow 4 (T + AOC) gave control over the algorithmic recommendations, both additionally enhancing behavioral agency. Lastly, Flow 5 (T + UDC + AOC) represented comprehensive empowerment, combining transparency with controls over both data and algorithmic outcomes to maximize agency. Figures \ref{fig:wireframe} and \ref{fig:sample} in Appendix \ref{appendix:figure} show the five-flow wireframe and representative finalized webpages.

Participants in all flows started with the same landing page informing the news recommender prototype. All five flows ended up displaying the same broadly appealing news snippets chosen from the pretest, as we focused more on users’ feelings on how the recommendations were made rather than the recommendation results themselves. This also controlled for bias due to content preferences and isolated the observed effects caused solely by the recommendation process. If declining cookies, participants still saw the same results but with a note stating that their data was not used and the recommendations were random. 

\subsubsection{Measurements} \label{measure}
Participants first indicated their interests in the selected news as a manipulation check to ensure the stimuli were engaging and relevant. To evaluate their experiences with the recommendation mechanism, we used items from \cite{pu2011user} to assess their perception of control (how much they think they can influence the system), as well as system explainability and transparency. Items from \cite{knijnenburg2012explaining} were used to assess the perception of system effectiveness (whether the system can help users make choices). We also included the human-computer trust scale (HCTS) \cite{gulati2019design} to measure users’ trust in the system. All questions were on a 5-point Likert scale (1 = Strongly disagree, 5 = Strongly agree).

\subsubsection{Data Analysis}
We conducted statistical analyses using R software, beginning by gaining a data overview (mean and SD) across conditions for each measurement. To examine the impact of the recommendation flows, we performed a Multivariate Analysis of Covariance (MANCOVA), controlling for demographics to address potential confounding effects and provide detailed context. Post-hoc analyses were then performed to pinpoint specific group differences where significant differences across conditions were observed, enabling a clear understanding of the condition effects.

\section{PRELIMINARY RESULTS}
As a basis for the evaluations participants provided, the stimuli manipulation check indicated that all participants found the news snippets relevant to their interest (\textit{M} = 3.53, \textit{SD} = 0.94). Table \ref{tab:mancova} displays the MANCOVA results, detailing the effects of the measurements and the covariates. The analysis identified significant differences across conditions for the perception of control toward the system (\textit{F} = 6.98, \textit{p} < .001).

Post-hoc results of perceived control from Table \ref{tab:posthoc} reveal that “Only T” scored significantly lower than conditions incorporating user control over data (T + UDC) (\textit{p} = .046), algorithm outcomes (T + AOC) (\textit{p} < .001), or both (T + UDC + AOC) (\textit{p} < .001). This variance suggests that information and power asymmetry are distinct phenomena, with users explicitly preferring actionable control beyond mere awareness. Interestingly, “Only T” also scored lower than the baseline “None” (\textit{p} = .008), implying that transparency awareness may backfire and be worse than having nothing. Additionally, “T + UDC”, as one type of control, scored lower than the dual-mechanism (T + UDC + AOC) (\textit{p} = .049), highlighting the added value of combining multiple control mechanisms.

\begin{table}[ht]
\setlength{\tabcolsep}{8pt}
\centering
\caption{MANCOVA Results (F-value) for Main Measurements and Covariates.}
\begin{tabular}{lccccc}
\hline
 & CON & EXPL & TRAN & EFFE & TRU \\
\hline
Condition & 6.98*** & 0.84 & 0.72 & 0.60 & 0.17 \\
Age       & 0.04 & 2.17 & 4.15* & 1.75 & 4.85* \\
Gender    & 0.52 & 0.23 & 0.00 & 0.47 & 0.19 \\
Education & 1.39 & 3.90* & 2.81 & 4.47* & 1.46 \\
Political & 0.81 & 0.92 & 0.56 & 0.63 & 0.11 \\
\hline
\end{tabular}
\parbox{\columnwidth}{\raggedright \footnotesize \textit{Note}: CON = Perceived Control; EXPL = System Explainability; TRAN = System Transparency; EFFE = System Effectiveness; TRU = Trust to System. *\textit{p} < .05, **\textit{p} < .01, ***\textit{p} < .001.}
\label{tab:mancova}
\end{table}

\begin{table}[ht]
\setlength{\tabcolsep}{6.8pt}
\centering
\caption{Post-hoc Analysis for the Perception of Control.}
\begin{tabular}{llllll}
\hline
 & None & \makecell{Only\\T} & \makecell{T +\\UDC} & \makecell{T +\\AOC} & \makecell{T +\\UDC\\+ AOC} \\
\hline
\makecell[l]{$\text{Mean}_{\text{est}}$} & 3.26 & 2.67 & 3.15 & 3.49 & 3.63 \\
\hline
\makecell[l]{$\text{Mean}_{\text{diff}}$} \\
\makecell[l]{\hspace{9pt}None} & & -0.59** & -0.10 & 0.24 & 0.38 \\
\makecell[l]{\hspace{9pt}Only T} & & & 0.48* & 0.83*** & 0.96*** \\
\makecell[l]{\hspace{10pt}T + UDC} & & & & 0.34 & 0.48* \\
\makecell[l]{\hspace{10pt}T + AOC} & & & & & 0.14 \\
\hline
\multicolumn{6}{l}{\footnotesize \textit{Note}: *\textit{p} < .05, **\textit{p} < .01, ***\textit{p} < .001.}
\end{tabular}
\label{tab:posthoc}
\end{table}

\section{DISCUSSION}

\subsection{Discussion and Interpretation of Results}
This study provides preliminary insights into the interplay between transparency, control, and user perceptions of agency in AI-driven environments. The marked improvement in perceived control from "Only T" to "T + AOC" highlights the potential and promising application of tools like the AOC slider in addressing power asymmetry and enhancing users' sense of behavioral agency (RQ1).

Moreover, while control over data collection (T + UDC) offered benefits over transparency alone, it did not perform as strongly as combining both data and outcome controls (T + UDC + AOC). This points out that addressing multiple dimensions of user control can better satisfy users’ needs. Interestingly, however, “T + AOC” performed comparably to the dual-control condition, which raises intriguing possibilities. It indicates that a direct and actionable control such as AOC may resonate more with users by offering immediate interaction with recommendation outputs, which may sufficiently fulfill their behaviorally agency expectations. In contrast, data control, as an invisible background operation, may feel more abstract. Additionally, with data control mechanisms being common in digital systems, users may view them as standard rather than empowering features, reducing their perceived value as a control. Another consideration is that the cognitive load caused by multiple control options may potentially dilute their combined effect. When an impactful control (e.g., AOC) is available, adding other less-targeted ones (e.g., UDC) may not necessarily yield additional benefits. Offering more options may also heighten users’ awareness of what is beyond their control, thereby offsetting any perceived gains in control. These findings emphasize the importance of striking a balance in designing user controls, ensuring they are direct, effective, yet not overwhelming to navigate.

The study also revealed differences in user perception and experience across types of agency (RQ2). We found that perceptual agency alone (Only T) led to significantly lower perceived control compared to conditions with behavioral agency, such as control over data (T + UDC), algorithm outcomes (T + AOC), or both (T + UDC + AOC). This highlights the limitations of perceptual agency in fostering a true sense of control. When users are informed about data collection without actionable ways to influence it, they may experience heightened feelings of powerlessness. Notably, “Only T” also scored lower in perceived control than the baseline "None", where no transparency or control cues were provided, further reflecting the counterproductive effect of perceptual agency without behavioral agency. In “None”, users may simply be unaware of the underlying algorithmic processes, resulting in a neutral perception of control. However, “Only T” made users acutely aware of data collection but denied them the ability to intervene, intensifying feelings of disempowerment. This pattern aligns with cognitive dissonance theory \cite{harmon2019introduction}, as the tension between being informed yet unable to act creates discomfort and lowers perceived agency. It also supports the need for user interventions during interaction with algorithmic systems in the human-AI synergy concept \cite{sundar2020rise}.

This study makes several contributions. First, it proposes a new idea of collaborative decision-making between humans and AI that allows users to control the degree of content recommended by AI—a step we claim is missing in current systems. While still exploratory, our experiments provided nuanced insights into user control, their interactions with AI systems, and their attitudes towards it. Second, this study is among the earliest to manipulate different types of agency in a comparative experimental setting. By examining these dynamics, we aim to promote an environment where users can develop a more positive and informed relationship with the technology they engage with.

\subsection{Design Implications}

\subsubsection{Coupling Transparency with Control in Algorithmic Systems} \label{control}
Our findings indicate that while transparency remains critical in ethical AI design, it must be coupled with corresponding controls to avoid paradoxically diminishing user experience. For example, beyond simply asking for acceptance or rejection of data usage, transparency regarding data collection can be enhanced by offering more options to modify the scope of data sharing or specify which types of data should influence algorithmic recommendations (e.g., adjust the weight given to demographic records or search history). More importantly, we provide promising insights into accompanying algorithm transparency with an active role for users in managing the recommendations they receive. This process can be enriched by adjusting the weight of recommendation type (e.g., tailored vs. serendipitous, based on demographic data vs. search history), setting special schedules (e.g., “serendipity time”), or even personalizing recommendation modes for distinct contexts (e.g., set proportions of AI-recommended content for “working/leisure” or “morning/night”).

\subsubsection{Balancing Algorithmic Efficiency and User Agency}
Our study reveals a key challenge–empowering users without overwhelming them with complexity. While transparency must go beyond simply exposing system processes to influence outcomes effectively, the collaborative relationship should be meaningful and effective. During this process, it is important to present users with clear and intuitive control mechanisms without requiring extensive learning. Designers should also ensure that control options do not burden users with too many numbers and types of choices. Moreover, given that different users prioritize different aspects of control, adaptive systems that tailor control mechanisms to individual preferences may enhance satisfaction and engagement. Some users may value granular control over algorithmic processes, while others may prefer simplified, high-level adjustments. Designing for flexibility ensures that diverse user needs are met without compromising usability.

\subsection{LIMITATION AND FUTURE PLAN}
Some imitations exist and open avenues for future research. First, significant differences were observed only in perceived control across conditions. While this supports the study as an initial proof of concept, the prototypes may not yet fully capture or address the complexities of user agency. Our AOC also lacked clarity on where the content came from outside the specified percentage, which may cause vagueness and uncertainty and limit deeper judgments like system effectiveness or trust. Future research could refine and expand the design to enhance performance on other measures. Second, this study operationalized agency mainly as perceived control–a key aspect of behavioral agency. However, agency is a broader construct encompassing multiple dimensions, and future work should incorporate a broader set of measures to comprehensively assess user agency. Third, while our quantitative approach provides valuable insights, it would be beneficial to explore the underlying reasons why users value the AOC slider and how it can be optimized for greater impact. Regarding the study stimuli, as a control of our study, the same 12 news snippets were presented regardless of participants' choices. While this deliberate design was to minimize content bias and isolate the effects of the recommendation flow, it may have impacted participants' perceptions of the system. Although we measured their interest in the recommended content, the average score was not extremely high, suggesting that each snippet may not have resonated with each participant. Incorrect or irrelevant recommendations could heighten privacy concerns and affect their judgments \cite{asthana2024know}. The relatively small number of 12 snippets may have also limited the perceived variety and depth of recommendations. Future design could explore how a more diverse and larger recommendation pool, or a dynamic system that adapts to user inputs, impacts their sense of control and engagement toward the system.

Moving forward, this preliminary research paves the way for the refinement of guidelines and designs for AI-based recommender systems. Beyond a simple AOC slider, we plan to expand on these initial findings by exploring and testing additional mechanisms regarding user agency. As outlined in Section \ref{control}, one key direction involves shifting more power to users in shaping their recommendation outputs; there is also ample room for innovations in personalization designs for diverse content discovery. To deepen our understanding, for the next step, we will incorporate qualitative methods, such as usability studies and observational research, to present and evaluate our ideas. These methods will allow us to gather direct feedback on user experiences regarding these new designs, further refining our prototypes to better meet user needs.

Despite the limitations, this study serves as an initial proof of concept and sets the stage for future research to expand the application and effectiveness of options that enhance user agency. By advancing this line of inquiry, we hope to contribute to the development of more ethical, user-centered AI systems and inspire new perspectives in HCI design.



\bibliographystyle{ACM-Reference-Format}
\bibliography{sample-base}


\begin{thebibliography}{54}


\ifx \showCODEN    \undefined \def \showCODEN     #1{\unskip}     \fi
\ifx \showISBNx    \undefined \def \showISBNx     #1{\unskip}     \fi
\ifx \showISBNxiii \undefined \def \showISBNxiii  #1{\unskip}     \fi
\ifx \showISSN     \undefined \def \showISSN      #1{\unskip}     \fi
\ifx \showLCCN     \undefined \def \showLCCN      #1{\unskip}     \fi
\ifx \shownote     \undefined \def \shownote      #1{#1}          \fi
\ifx \showarticletitle \undefined \def \showarticletitle #1{#1}   \fi
\ifx \showURL      \undefined \def \showURL       {\relax}        \fi
\providecommand\bibfield[2]{#2}
\providecommand\bibinfo[2]{#2}
\providecommand\natexlab[1]{#1}
\providecommand\showeprint[2][]{arXiv:#2}

\bibitem[Adamopoulos and Tuzhilin(2014)]%
        {adamopoulos2014over}
\bibfield{author}{\bibinfo{person}{Panagiotis Adamopoulos} {and} \bibinfo{person}{Alexander Tuzhilin}.} \bibinfo{year}{2014}\natexlab{}.
\newblock \showarticletitle{On over-specialization and concentration bias of recommendations: Probabilistic neighborhood selection in collaborative filtering systems}. In \bibinfo{booktitle}{\emph{Proceedings of the 8th ACM Conference on Recommender systems}}. \bibinfo{pages}{153--160}.
\newblock


\bibitem[Amershi et~al\mbox{.}(2019)]%
        {amershi2019guidelines}
\bibfield{author}{\bibinfo{person}{Saleema Amershi}, \bibinfo{person}{Dan Weld}, \bibinfo{person}{Mihaela Vorvoreanu}, \bibinfo{person}{Adam Fourney}, \bibinfo{person}{Besmira Nushi}, \bibinfo{person}{Penny Collisson}, \bibinfo{person}{Jina Suh}, \bibinfo{person}{Shamsi Iqbal}, \bibinfo{person}{Paul~N Bennett}, \bibinfo{person}{Kori Inkpen}, {et~al\mbox{.}}} \bibinfo{year}{2019}\natexlab{}.
\newblock \showarticletitle{Guidelines for human-AI interaction}. In \bibinfo{booktitle}{\emph{Proceedings of the 2019 chi conference on human factors in computing systems}}. \bibinfo{pages}{1--13}.
\newblock


\bibitem[Asthana et~al\mbox{.}(2024)]%
        {asthana2024know}
\bibfield{author}{\bibinfo{person}{Sumit Asthana}, \bibinfo{person}{Jane Im}, \bibinfo{person}{Zhe Chen}, {and} \bibinfo{person}{Nikola Banovic}.} \bibinfo{year}{2024}\natexlab{}.
\newblock \showarticletitle{" I know even if you don't tell me": Understanding Users' Privacy Preferences Regarding AI-based Inferences of Sensitive Information for Personalization}. In \bibinfo{booktitle}{\emph{Proceedings of the 2024 CHI Conference on Human Factors in Computing Systems}}. \bibinfo{pages}{1--21}.
\newblock


\bibitem[Bakalov et~al\mbox{.}(2013)]%
        {bakalov2013approach}
\bibfield{author}{\bibinfo{person}{Fedor Bakalov}, \bibinfo{person}{Marie-Jean Meurs}, \bibinfo{person}{Birgitta K{\"o}nig-Ries}, \bibinfo{person}{Bahar Sateli}, \bibinfo{person}{Ren{\'e} Witte}, \bibinfo{person}{Greg Butler}, {and} \bibinfo{person}{Adrian Tsang}.} \bibinfo{year}{2013}\natexlab{}.
\newblock \showarticletitle{An approach to controlling user models and personalization effects in recommender systems}. In \bibinfo{booktitle}{\emph{Proceedings of the 2013 international conference on Intelligent user interfaces}}. \bibinfo{pages}{49--56}.
\newblock


\bibitem[Berens et~al\mbox{.}(2024)]%
        {berens2024cookie}
\bibfield{author}{\bibinfo{person}{Benjamin~Maximilian Berens}, \bibinfo{person}{Mark Bohlender}, \bibinfo{person}{Heike Dietmann}, \bibinfo{person}{Chiara Krisam}, \bibinfo{person}{Oksana Kulyk}, {and} \bibinfo{person}{Melanie Volkamer}.} \bibinfo{year}{2024}\natexlab{}.
\newblock \showarticletitle{Cookie disclaimers: Dark patterns and lack of transparency}.
\newblock \bibinfo{journal}{\emph{Computers \& Security}}  \bibinfo{volume}{136} (\bibinfo{year}{2024}), \bibinfo{pages}{103507}.
\newblock


\bibitem[Burrell(2016)]%
        {burrell2016machine}
\bibfield{author}{\bibinfo{person}{Jenna Burrell}.} \bibinfo{year}{2016}\natexlab{}.
\newblock \showarticletitle{How the machine ‘thinks’: Understanding opacity in machine learning algorithms}.
\newblock \bibinfo{journal}{\emph{Big data \& society}} \bibinfo{volume}{3}, \bibinfo{number}{1} (\bibinfo{year}{2016}), \bibinfo{pages}{2053951715622512}.
\newblock


\bibitem[Chen and Sundar(2018)]%
        {chen2018app}
\bibfield{author}{\bibinfo{person}{Tsai-Wei Chen} {and} \bibinfo{person}{S~Shyam Sundar}.} \bibinfo{year}{2018}\natexlab{}.
\newblock \showarticletitle{This app would like to use your current location to better serve you: Importance of user assent and system transparency in personalized mobile services}. In \bibinfo{booktitle}{\emph{Proceedings of the 2018 CHI Conference on Human Factors in Computing Systems}}. \bibinfo{publisher}{ACM}, \bibinfo{address}{USA}, \bibinfo{pages}{1--13}.
\newblock


\bibitem[Cheng and Chouldechova(2023)]%
        {cheng2023overcoming}
\bibfield{author}{\bibinfo{person}{Lingwei Cheng} {and} \bibinfo{person}{Alexandra Chouldechova}.} \bibinfo{year}{2023}\natexlab{}.
\newblock \showarticletitle{Overcoming Algorithm Aversion: A Comparison between Process and Outcome Control}. In \bibinfo{booktitle}{\emph{Proceedings of the 2023 CHI Conference on Human Factors in Computing Systems}}. \bibinfo{publisher}{ACM}, \bibinfo{address}{USA}, \bibinfo{pages}{1--27}.
\newblock


\bibitem[Conover et~al\mbox{.}(2011)]%
        {conover2011political}
\bibfield{author}{\bibinfo{person}{Michael Conover}, \bibinfo{person}{Jacob Ratkiewicz}, \bibinfo{person}{Matthew Francisco}, \bibinfo{person}{Bruno Gon{\c{c}}alves}, \bibinfo{person}{Filippo Menczer}, {and} \bibinfo{person}{Alessandro Flammini}.} \bibinfo{year}{2011}\natexlab{}.
\newblock \showarticletitle{Political polarization on twitter}. In \bibinfo{booktitle}{\emph{Proceedings of the International AAAI Conference on Web and Social Media}}, Vol.~\bibinfo{volume}{5-1}. \bibinfo{publisher}{AAAI}, \bibinfo{address}{USA}, \bibinfo{pages}{89--96}.
\newblock


\bibitem[Coyle et~al\mbox{.}(2012)]%
        {coyle2012did}
\bibfield{author}{\bibinfo{person}{David Coyle}, \bibinfo{person}{James Moore}, \bibinfo{person}{Per~Ola Kristensson}, \bibinfo{person}{Paul Fletcher}, {and} \bibinfo{person}{Alan Blackwell}.} \bibinfo{year}{2012}\natexlab{}.
\newblock \showarticletitle{I did that! Measuring users' experience of agency in their own actions}. In \bibinfo{booktitle}{\emph{Proceedings of the SIGCHI conference on human factors in computing systems}}. \bibinfo{pages}{2025--2034}.
\newblock


\bibitem[Cramer et~al\mbox{.}(2008)]%
        {cramer2008effects}
\bibfield{author}{\bibinfo{person}{Henriette Cramer}, \bibinfo{person}{Vanessa Evers}, \bibinfo{person}{Satyan Ramlal}, \bibinfo{person}{Maarten Van~Someren}, \bibinfo{person}{Lloyd Rutledge}, \bibinfo{person}{Natalia Stash}, \bibinfo{person}{Lora Aroyo}, {and} \bibinfo{person}{Bob Wielinga}.} \bibinfo{year}{2008}\natexlab{}.
\newblock \showarticletitle{The effects of transparency on trust in and acceptance of a content-based art recommender}.
\newblock \bibinfo{journal}{\emph{User Modeling and User-adapted Interaction}}  \bibinfo{volume}{18} (\bibinfo{year}{2008}), \bibinfo{pages}{455--496}.
\newblock


\bibitem[Desantis et~al\mbox{.}(2016)]%
        {desantis2016agency}
\bibfield{author}{\bibinfo{person}{Andrea Desantis}, \bibinfo{person}{Florian Waszak}, {and} \bibinfo{person}{Andrei Gorea}.} \bibinfo{year}{2016}\natexlab{}.
\newblock \showarticletitle{Agency alters perceptual decisions about action-outcomes}.
\newblock \bibinfo{journal}{\emph{Experimental brain research}}  \bibinfo{volume}{234} (\bibinfo{year}{2016}), \bibinfo{pages}{2819--2827}.
\newblock


\bibitem[Ehsan et~al\mbox{.}(2021)]%
        {ehsan2021operationalizing}
\bibfield{author}{\bibinfo{person}{Upol Ehsan}, \bibinfo{person}{Philipp Wintersberger}, \bibinfo{person}{Q~Vera Liao}, \bibinfo{person}{Martina Mara}, \bibinfo{person}{Marc Streit}, \bibinfo{person}{Sandra Wachter}, \bibinfo{person}{Andreas Riener}, {and} \bibinfo{person}{Mark~O Riedl}.} \bibinfo{year}{2021}\natexlab{}.
\newblock \showarticletitle{Operationalizing human-centered perspectives in explainable AI}. In \bibinfo{booktitle}{\emph{Extended abstracts of the 2021 CHI conference on human factors in computing systems}}. \bibinfo{pages}{1--6}.
\newblock


\bibitem[Eiband et~al\mbox{.}(2018)]%
        {eiband2018bringing}
\bibfield{author}{\bibinfo{person}{Malin Eiband}, \bibinfo{person}{Hanna Schneider}, \bibinfo{person}{Mark Bilandzic}, \bibinfo{person}{Julian Fazekas-Con}, \bibinfo{person}{Mareike Haug}, {and} \bibinfo{person}{Heinrich Hussmann}.} \bibinfo{year}{2018}\natexlab{}.
\newblock \showarticletitle{Bringing transparency design into practice}. In \bibinfo{booktitle}{\emph{23rd international conference on intelligent user interfaces}}. \bibinfo{publisher}{ACM}, \bibinfo{address}{USA}, \bibinfo{pages}{211--223}.
\newblock


\bibitem[Endsley(2023)]%
        {endsley2023supporting}
\bibfield{author}{\bibinfo{person}{Mica~R Endsley}.} \bibinfo{year}{2023}\natexlab{}.
\newblock \showarticletitle{Supporting Human-AI Teams: Transparency, explainability, and situation awareness}.
\newblock \bibinfo{journal}{\emph{Computers in Human Behavior}}  \bibinfo{volume}{140} (\bibinfo{year}{2023}), \bibinfo{pages}{107574}.
\newblock


\bibitem[Felzmann et~al\mbox{.}(2020)]%
        {felzmann2020towards}
\bibfield{author}{\bibinfo{person}{Heike Felzmann}, \bibinfo{person}{Eduard Fosch-Villaronga}, \bibinfo{person}{Christoph Lutz}, {and} \bibinfo{person}{Aurelia Tam{\`o}-Larrieux}.} \bibinfo{year}{2020}\natexlab{}.
\newblock \showarticletitle{Towards transparency by design for artificial intelligence}.
\newblock \bibinfo{journal}{\emph{Science and Engineering Ethics}} \bibinfo{volume}{26}, \bibinfo{number}{6} (\bibinfo{year}{2020}), \bibinfo{pages}{3333--3361}.
\newblock


\bibitem[Gulati et~al\mbox{.}(2019)]%
        {gulati2019design}
\bibfield{author}{\bibinfo{person}{Siddharth Gulati}, \bibinfo{person}{Sonia Sousa}, {and} \bibinfo{person}{David Lamas}.} \bibinfo{year}{2019}\natexlab{}.
\newblock \showarticletitle{Design, development and evaluation of a human-computer trust scale}.
\newblock \bibinfo{journal}{\emph{Behaviour \& Information Technology}} \bibinfo{volume}{38}, \bibinfo{number}{10} (\bibinfo{year}{2019}), \bibinfo{pages}{1004--1015}.
\newblock


\bibitem[Harambam et~al\mbox{.}(2019)]%
        {harambam2019designing}
\bibfield{author}{\bibinfo{person}{Jaron Harambam}, \bibinfo{person}{Dimitrios Bountouridis}, \bibinfo{person}{Mykola Makhortykh}, {and} \bibinfo{person}{Joris Van~Hoboken}.} \bibinfo{year}{2019}\natexlab{}.
\newblock \showarticletitle{Designing for the better by taking users into account: A qualitative evaluation of user control mechanisms in (news) recommender systems}. In \bibinfo{booktitle}{\emph{Proceedings of the 13th ACM conference on recommender systems}}. \bibinfo{pages}{69--77}.
\newblock


\bibitem[Harmon-Jones and Mills(2019)]%
        {harmon2019introduction}
\bibfield{author}{\bibinfo{person}{Eddie Harmon-Jones} {and} \bibinfo{person}{Judson Mills}.} \bibinfo{year}{2019}\natexlab{}.
\newblock \showarticletitle{An introduction to cognitive dissonance theory and an overview of current perspectives on the theory.}
\newblock \bibinfo{journal}{\emph{American Psychological Association}} (\bibinfo{year}{2019}).
\newblock


\bibitem[Harper et~al\mbox{.}(2015)]%
        {harper2015putting}
\bibfield{author}{\bibinfo{person}{F~Maxwell Harper}, \bibinfo{person}{Funing Xu}, \bibinfo{person}{Harmanpreet Kaur}, \bibinfo{person}{Kyle Condiff}, \bibinfo{person}{Shuo Chang}, {and} \bibinfo{person}{Loren Terveen}.} \bibinfo{year}{2015}\natexlab{}.
\newblock \showarticletitle{Putting users in control of their recommendations}. In \bibinfo{booktitle}{\emph{Proceedings of the 9th ACM Conference on Recommender Systems}}. \bibinfo{pages}{3--10}.
\newblock


\bibitem[Jannach et~al\mbox{.}(2019)]%
        {jannach2019explanations}
\bibfield{author}{\bibinfo{person}{Dietmar Jannach}, \bibinfo{person}{Michael Jugovac}, {and} \bibinfo{person}{Ingrid Nunes}.} \bibinfo{year}{2019}\natexlab{}.
\newblock \showarticletitle{Explanations and user control in recommender systems}. In \bibinfo{booktitle}{\emph{Proceedings of the 23rd International Workshop on Personalization and Recommendation on the Web and Beyond}}. \bibinfo{pages}{31--31}.
\newblock


\bibitem[Jannach et~al\mbox{.}(2017)]%
        {jannach2017user}
\bibfield{author}{\bibinfo{person}{Dietmar Jannach}, \bibinfo{person}{Sidra Naveed}, {and} \bibinfo{person}{Michael Jugovac}.} \bibinfo{year}{2017}\natexlab{}.
\newblock \showarticletitle{User control in recommender systems: Overview and interaction challenges}. In \bibinfo{booktitle}{\emph{E-Commerce and Web Technologies: 17th International Conference, EC-Web 2016, Porto, Portugal, September 5-8, 2016, Revised Selected Papers 17}}. Springer, \bibinfo{pages}{21--33}.
\newblock


\bibitem[Jin et~al\mbox{.}(2021)]%
        {jin2021key}
\bibfield{author}{\bibinfo{person}{Yucheng Jin}, \bibinfo{person}{Li Chen}, \bibinfo{person}{Wanling Cai}, {and} \bibinfo{person}{Pearl Pu}.} \bibinfo{year}{2021}\natexlab{}.
\newblock \showarticletitle{Key qualities of conversational recommender systems: From users’ perspective}. In \bibinfo{booktitle}{\emph{Proceedings of the 9th International Conference on Human-Agent Interaction}}. \bibinfo{pages}{93--102}.
\newblock


\bibitem[Jin et~al\mbox{.}(2020)]%
        {jin2020effects}
\bibfield{author}{\bibinfo{person}{Yucheng Jin}, \bibinfo{person}{Nava Tintarev}, \bibinfo{person}{Nyi~Nyi Htun}, {and} \bibinfo{person}{Katrien Verbert}.} \bibinfo{year}{2020}\natexlab{}.
\newblock \showarticletitle{Effects of personal characteristics in control-oriented user interfaces for music recommender systems}.
\newblock \bibinfo{journal}{\emph{User Modeling and User-Adapted Interaction}} \bibinfo{volume}{30}, \bibinfo{number}{2} (\bibinfo{year}{2020}), \bibinfo{pages}{199--249}.
\newblock


\bibitem[Kaminskas and Bridge(2016)]%
        {kaminskas2016diversity}
\bibfield{author}{\bibinfo{person}{Marius Kaminskas} {and} \bibinfo{person}{Derek Bridge}.} \bibinfo{year}{2016}\natexlab{}.
\newblock \showarticletitle{Diversity, serendipity, novelty, and coverage: a survey and empirical analysis of beyond-accuracy objectives in recommender systems}.
\newblock \bibinfo{journal}{\emph{ACM Transactions on Interactive Intelligent Systems (TiiS)}} \bibinfo{volume}{7}, \bibinfo{number}{1} (\bibinfo{year}{2016}), \bibinfo{pages}{1--42}.
\newblock


\bibitem[Kizilcec(2016)]%
        {kizilcec2016much}
\bibfield{author}{\bibinfo{person}{Ren{\'e}~F Kizilcec}.} \bibinfo{year}{2016}\natexlab{}.
\newblock \showarticletitle{How much information? Effects of transparency on trust in an algorithmic interface}. In \bibinfo{booktitle}{\emph{Proceedings of the 2016 CHI Conference on Human Factors in Computing Systems}}. \bibinfo{publisher}{ACM}, \bibinfo{address}{USA}, \bibinfo{pages}{2390--2395}.
\newblock


\bibitem[Knijnenburg et~al\mbox{.}(2012)]%
        {knijnenburg2012explaining}
\bibfield{author}{\bibinfo{person}{Bart~P Knijnenburg}, \bibinfo{person}{Martijn~C Willemsen}, \bibinfo{person}{Zeno Gantner}, \bibinfo{person}{Hakan Soncu}, {and} \bibinfo{person}{Chris Newell}.} \bibinfo{year}{2012}\natexlab{}.
\newblock \showarticletitle{Explaining the user experience of recommender systems}.
\newblock \bibinfo{journal}{\emph{User modeling and user-adapted interaction}}  \bibinfo{volume}{22} (\bibinfo{year}{2012}), \bibinfo{pages}{441--504}.
\newblock


\bibitem[Laban and Araujo(2020)]%
        {laban2020effect}
\bibfield{author}{\bibinfo{person}{Guy Laban} {and} \bibinfo{person}{Theo Araujo}.} \bibinfo{year}{2020}\natexlab{}.
\newblock \showarticletitle{The effect of personalization techniques in users' perceptions of conversational recommender systems}. In \bibinfo{booktitle}{\emph{Proceedings of the 20th ACM international conference on intelligent virtual agents}}. \bibinfo{pages}{1--3}.
\newblock


\bibitem[Lai et~al\mbox{.}(2023)]%
        {lai2023towards}
\bibfield{author}{\bibinfo{person}{Vivian Lai}, \bibinfo{person}{Chacha Chen}, \bibinfo{person}{Alison Smith-Renner}, \bibinfo{person}{Q~Vera Liao}, {and} \bibinfo{person}{Chenhao Tan}.} \bibinfo{year}{2023}\natexlab{}.
\newblock \showarticletitle{Towards a Science of Human-AI Decision Making: An Overview of Design Space in Empirical Human-Subject Studies}. In \bibinfo{booktitle}{\emph{Proceedings of the 2023 ACM Conference on Fairness, Accountability, and Transparency}}. \bibinfo{publisher}{ACM}, \bibinfo{address}{USA}, \bibinfo{pages}{1369--1385}.
\newblock


\bibitem[Liao et~al\mbox{.}(2020)]%
        {liao2020questioning}
\bibfield{author}{\bibinfo{person}{Q~Vera Liao}, \bibinfo{person}{Daniel Gruen}, {and} \bibinfo{person}{Sarah Miller}.} \bibinfo{year}{2020}\natexlab{}.
\newblock \showarticletitle{Questioning the AI: informing design practices for explainable AI user experiences}. In \bibinfo{booktitle}{\emph{Proceedings of the 2020 CHI conference on human factors in computing systems}}. \bibinfo{pages}{1--15}.
\newblock


\bibitem[Millett et~al\mbox{.}(2001)]%
        {millett2001cookies}
\bibfield{author}{\bibinfo{person}{Lynette~I Millett}, \bibinfo{person}{Batya Friedman}, {and} \bibinfo{person}{Edward Felten}.} \bibinfo{year}{2001}\natexlab{}.
\newblock \showarticletitle{Cookies and web browser design: Toward realizing informed consent online}. In \bibinfo{booktitle}{\emph{Proceedings of the SIGCHI conference on Human factors in computing systems}}. \bibinfo{publisher}{ACM}, \bibinfo{address}{USA}, \bibinfo{pages}{46--52}.
\newblock


\bibitem[Molina and Sundar(2022)]%
        {molina2022ai}
\bibfield{author}{\bibinfo{person}{Maria~D Molina} {and} \bibinfo{person}{S~Shyam Sundar}.} \bibinfo{year}{2022}\natexlab{}.
\newblock \showarticletitle{When AI moderates online content: effects of human collaboration and interactive transparency on user trust}.
\newblock \bibinfo{journal}{\emph{Journal of Computer-Mediated Communication}} \bibinfo{volume}{27}, \bibinfo{number}{4} (\bibinfo{year}{2022}), \bibinfo{pages}{zmac010}.
\newblock


\bibitem[Nguyen et~al\mbox{.}(2014)]%
        {nguyen2014exploring}
\bibfield{author}{\bibinfo{person}{Tien~T Nguyen}, \bibinfo{person}{Pik-Mai Hui}, \bibinfo{person}{F~Maxwell Harper}, \bibinfo{person}{Loren Terveen}, {and} \bibinfo{person}{Joseph~A Konstan}.} \bibinfo{year}{2014}\natexlab{}.
\newblock \showarticletitle{Exploring the filter bubble: the effect of using recommender systems on content diversity}. In \bibinfo{booktitle}{\emph{Proceedings of the 23rd International Conference on World Wide Web}}. \bibinfo{publisher}{ACM}, \bibinfo{address}{USA}, \bibinfo{pages}{677--686}.
\newblock


\bibitem[Pandey et~al\mbox{.}(2011)]%
        {pandey2011learning}
\bibfield{author}{\bibinfo{person}{Sandeep Pandey}, \bibinfo{person}{Mohamed Aly}, \bibinfo{person}{Abraham Bagherjeiran}, \bibinfo{person}{Andrew Hatch}, \bibinfo{person}{Peter Ciccolo}, \bibinfo{person}{Adwait Ratnaparkhi}, {and} \bibinfo{person}{Martin Zinkevich}.} \bibinfo{year}{2011}\natexlab{}.
\newblock \showarticletitle{Learning to target: what works for behavioral targeting}. In \bibinfo{booktitle}{\emph{Proceedings of the 20th ACM international conference on Information and knowledge management}}. \bibinfo{pages}{1805--1814}.
\newblock


\bibitem[Pasquale(2015)]%
        {pasquale2015black}
\bibfield{author}{\bibinfo{person}{Frank Pasquale}.} \bibinfo{year}{2015}\natexlab{}.
\newblock \bibinfo{booktitle}{\emph{The black box society: The secret algorithms that control money and information}}.
\newblock \bibinfo{publisher}{Harvard University Press}, \bibinfo{address}{USA}.
\newblock


\bibitem[Pu et~al\mbox{.}(2011)]%
        {pu2011user}
\bibfield{author}{\bibinfo{person}{Pearl Pu}, \bibinfo{person}{Li Chen}, {and} \bibinfo{person}{Rong Hu}.} \bibinfo{year}{2011}\natexlab{}.
\newblock \showarticletitle{A user-centric evaluation framework for recommender systems}. In \bibinfo{booktitle}{\emph{Proceedings of the fifth ACM conference on Recommender systems}}. \bibinfo{publisher}{ACM}, \bibinfo{address}{USA}, \bibinfo{pages}{157--164}.
\newblock


\bibitem[Rader et~al\mbox{.}(2018)]%
        {rader2018explanations}
\bibfield{author}{\bibinfo{person}{Emilee Rader}, \bibinfo{person}{Kelley Cotter}, {and} \bibinfo{person}{Janghee Cho}.} \bibinfo{year}{2018}\natexlab{}.
\newblock \showarticletitle{Explanations as mechanisms for supporting algorithmic transparency}. In \bibinfo{booktitle}{\emph{Proceedings of the 2018 CHI Conference on Human Factors in Computing Systems}}. \bibinfo{publisher}{ACM}, \bibinfo{address}{USA}, \bibinfo{pages}{1--13}.
\newblock


\bibitem[Ras et~al\mbox{.}(2018)]%
        {ras2018explanation}
\bibfield{author}{\bibinfo{person}{Gabri{\"e}lle Ras}, \bibinfo{person}{Marcel van Gerven}, {and} \bibinfo{person}{Pim Haselager}.} \bibinfo{year}{2018}\natexlab{}.
\newblock \showarticletitle{Explanation methods in deep learning: Users, values, concerns and challenges}.
\newblock \bibinfo{journal}{\emph{Explainable and interpretable models in computer vision and machine learning}} (\bibinfo{year}{2018}), \bibinfo{pages}{19--36}.
\newblock


\bibitem[Rashid et~al\mbox{.}(2002)]%
        {rashid2002getting}
\bibfield{author}{\bibinfo{person}{Al~Mamunur Rashid}, \bibinfo{person}{Istvan Albert}, \bibinfo{person}{Dan Cosley}, \bibinfo{person}{Shyong~K Lam}, \bibinfo{person}{Sean~M McNee}, \bibinfo{person}{Joseph~A Konstan}, {and} \bibinfo{person}{John Riedl}.} \bibinfo{year}{2002}\natexlab{}.
\newblock \showarticletitle{Getting to know you: learning new user preferences in recommender systems}. In \bibinfo{booktitle}{\emph{Proceedings of the 7th international conference on Intelligent user interfaces}}. \bibinfo{publisher}{ACM}, \bibinfo{address}{USA}, \bibinfo{pages}{127--134}.
\newblock


\bibitem[Rowland(2011)]%
        {rowland2011filter}
\bibfield{author}{\bibinfo{person}{Fred Rowland}.} \bibinfo{year}{2011}\natexlab{}.
\newblock \showarticletitle{The filter bubble: what the internet is hiding from you}.
\newblock \bibinfo{journal}{\emph{portal: Libraries and the Academy}} \bibinfo{volume}{11}, \bibinfo{number}{4} (\bibinfo{year}{2011}), \bibinfo{pages}{1009--1011}.
\newblock


\bibitem[R{\"u}hr et~al\mbox{.}(2023)]%
        {ruhr2023intelligent}
\bibfield{author}{\bibinfo{person}{Alexander R{\"u}hr}, \bibinfo{person}{Benedikt Berger}, {and} \bibinfo{person}{Thomas Hess}.} \bibinfo{year}{2023}\natexlab{}.
\newblock \showarticletitle{Intelligent IT Systems in Business Application: Control and Transparency as Means of Building Trust in AI}.
\newblock In \bibinfo{booktitle}{\emph{Work and AI 2030: Challenges and Strategies for Tomorrow's Work}}. \bibinfo{publisher}{Springer}, \bibinfo{address}{USA}, \bibinfo{pages}{125--132}.
\newblock


\bibitem[Schafer et~al\mbox{.}(2002)]%
        {schafer2002meta}
\bibfield{author}{\bibinfo{person}{J~Ben Schafer}, \bibinfo{person}{Joseph~A Konstan}, {and} \bibinfo{person}{John Riedl}.} \bibinfo{year}{2002}\natexlab{}.
\newblock \showarticletitle{Meta-recommendation systems: user-controlled integration of diverse recommendations}. In \bibinfo{booktitle}{\emph{Proceedings of the eleventh international conference on Information and knowledge management}}. \bibinfo{pages}{43--51}.
\newblock


\bibitem[Schemmer et~al\mbox{.}(2022)]%
        {schemmer2022meta}
\bibfield{author}{\bibinfo{person}{Max Schemmer}, \bibinfo{person}{Patrick Hemmer}, \bibinfo{person}{Maximilian Nitsche}, \bibinfo{person}{Niklas K{\"u}hl}, {and} \bibinfo{person}{Michael V{\"o}ssing}.} \bibinfo{year}{2022}\natexlab{}.
\newblock \showarticletitle{A meta-analysis of the utility of explainable artificial intelligence in human-AI decision-making}. In \bibinfo{booktitle}{\emph{Proceedings of the 2022 AAAI/ACM Conference on AI, Ethics, and Society}}. \bibinfo{publisher}{ACM}, \bibinfo{address}{USA}, \bibinfo{pages}{617--626}.
\newblock


\bibitem[Segijn et~al\mbox{.}(2021)]%
        {segijn2021literature}
\bibfield{author}{\bibinfo{person}{Claire~M Segijn}, \bibinfo{person}{Joanna Strycharz}, \bibinfo{person}{Amy Riegelman}, {and} \bibinfo{person}{Cody Hennesy}.} \bibinfo{year}{2021}\natexlab{}.
\newblock \showarticletitle{A literature review of personalization transparency and control: introducing the transparency--awareness--control Framework}.
\newblock \bibinfo{journal}{\emph{Media and Communication}} \bibinfo{volume}{9}, \bibinfo{number}{4} (\bibinfo{year}{2021}), \bibinfo{pages}{120--133}.
\newblock


\bibitem[Shin and Park(2019)]%
        {shin2019role}
\bibfield{author}{\bibinfo{person}{Donghee Shin} {and} \bibinfo{person}{Yong~Jin Park}.} \bibinfo{year}{2019}\natexlab{}.
\newblock \showarticletitle{Role of fairness, accountability, and transparency in algorithmic affordance}.
\newblock \bibinfo{journal}{\emph{Computers in Human Behavior}}  \bibinfo{volume}{98} (\bibinfo{year}{2019}), \bibinfo{pages}{277--284}.
\newblock


\bibitem[Sonboli et~al\mbox{.}(2021)]%
        {sonboli2021fairness}
\bibfield{author}{\bibinfo{person}{Nasim Sonboli}, \bibinfo{person}{Jessie~J Smith}, \bibinfo{person}{Florencia Cabral~Berenfus}, \bibinfo{person}{Robin Burke}, {and} \bibinfo{person}{Casey Fiesler}.} \bibinfo{year}{2021}\natexlab{}.
\newblock \showarticletitle{Fairness and transparency in recommendation: The users’ perspective}. In \bibinfo{booktitle}{\emph{Proceedings of the 29th ACM Conference on User Modeling, Adaptation and Personalization}}. \bibinfo{publisher}{ACM}, \bibinfo{address}{USA}, \bibinfo{pages}{274--279}.
\newblock


\bibitem[Sundar(2020)]%
        {sundar2020rise}
\bibfield{author}{\bibinfo{person}{S~Shyam Sundar}.} \bibinfo{year}{2020}\natexlab{}.
\newblock \showarticletitle{Rise of machine agency: A framework for studying the psychology of human-AI interaction (HAII)}.
\newblock \bibinfo{journal}{\emph{Journal of Computer-Mediated Communication}} \bibinfo{volume}{25}, \bibinfo{number}{1} (\bibinfo{year}{2020}), \bibinfo{pages}{74--88}.
\newblock


\bibitem[Tsai and Brusilovsky(2021)]%
        {tsai2021effects}
\bibfield{author}{\bibinfo{person}{Chun-Hua Tsai} {and} \bibinfo{person}{Peter Brusilovsky}.} \bibinfo{year}{2021}\natexlab{}.
\newblock \showarticletitle{The effects of controllability and explainability in a social recommender system}.
\newblock \bibinfo{journal}{\emph{User Modeling and User-Adapted Interaction}}  \bibinfo{volume}{31} (\bibinfo{year}{2021}), \bibinfo{pages}{591--627}.
\newblock


\bibitem[Weller(2019)]%
        {weller2019transparency}
\bibfield{author}{\bibinfo{person}{Adrian Weller}.} \bibinfo{year}{2019}\natexlab{}.
\newblock \showarticletitle{Transparency: motivations and challenges}.
\newblock In \bibinfo{booktitle}{\emph{Explainable AI: Interpreting, Explaining and Visualizing Deep Learning}}. \bibinfo{publisher}{Springer}, \bibinfo{address}{USA}, \bibinfo{pages}{23--40}.
\newblock


\bibitem[Wolf(2019)]%
        {wolf2019explainability}
\bibfield{author}{\bibinfo{person}{Christine~T Wolf}.} \bibinfo{year}{2019}\natexlab{}.
\newblock \showarticletitle{Explainability scenarios: towards scenario-based XAI design}. In \bibinfo{booktitle}{\emph{Proceedings of the 24th International Conference on Intelligent User Interfaces}}. \bibinfo{publisher}{ACM}, \bibinfo{address}{USA}, \bibinfo{pages}{252--257}.
\newblock


\bibitem[Yoneda et~al\mbox{.}(2019)]%
        {yoneda2019algorithms}
\bibfield{author}{\bibinfo{person}{Takeshi Yoneda}, \bibinfo{person}{Shunsuke Kozawa}, \bibinfo{person}{Keisuke Osone}, \bibinfo{person}{Yukinori Koide}, \bibinfo{person}{Yosuke Abe}, {and} \bibinfo{person}{Yoshifumi Seki}.} \bibinfo{year}{2019}\natexlab{}.
\newblock \showarticletitle{Algorithms and system architecture for immediate personalized news recommendations}. In \bibinfo{booktitle}{\emph{IEEE/WIC/ACM International Conference on Web Intelligence}}. \bibinfo{pages}{124--131}.
\newblock


\bibitem[Zhang and Sundar(2019)]%
        {zhang2019proactive}
\bibfield{author}{\bibinfo{person}{Bo Zhang} {and} \bibinfo{person}{S~Shyam Sundar}.} \bibinfo{year}{2019}\natexlab{}.
\newblock \showarticletitle{Proactive vs. reactive personalization: Can customization of privacy enhance user experience?}
\newblock \bibinfo{journal}{\emph{International journal of human-computer studies}}  \bibinfo{volume}{128} (\bibinfo{year}{2019}), \bibinfo{pages}{86--99}.
\newblock


\bibitem[Zheng et~al\mbox{.}(2022)]%
        {zheng2022perd}
\bibfield{author}{\bibinfo{person}{Xuanzhi Zheng}, \bibinfo{person}{Guoshuai Zhao}, \bibinfo{person}{Li Zhu}, {and} \bibinfo{person}{Xueming Qian}.} \bibinfo{year}{2022}\natexlab{}.
\newblock \showarticletitle{PERD: Personalized emoji recommendation with dynamic user preference}. In \bibinfo{booktitle}{\emph{Proceedings of the 45th international ACM SIGIR conference on research and development in information retrieval}}. \bibinfo{publisher}{ACM}, \bibinfo{address}{USA}, \bibinfo{pages}{1922--1926}.
\newblock


\bibitem[Ziarani and Ravanmehr(2021)]%
        {ziarani2021serendipity}
\bibfield{author}{\bibinfo{person}{Reza~Jafari Ziarani} {and} \bibinfo{person}{Reza Ravanmehr}.} \bibinfo{year}{2021}\natexlab{}.
\newblock \showarticletitle{Serendipity in recommender systems: a systematic literature review}.
\newblock \bibinfo{journal}{\emph{Journal of Computer Science and Technology}}  \bibinfo{volume}{36} (\bibinfo{year}{2021}), \bibinfo{pages}{375--396}.
\newblock


\end{thebibliography}

\newpage
\appendix
\onecolumn
\section{Appendix: Prototype Figures}
\label{appendix:figure}

\begin{figure*}[ht]
    \centering
    \includegraphics[width=0.943\linewidth]{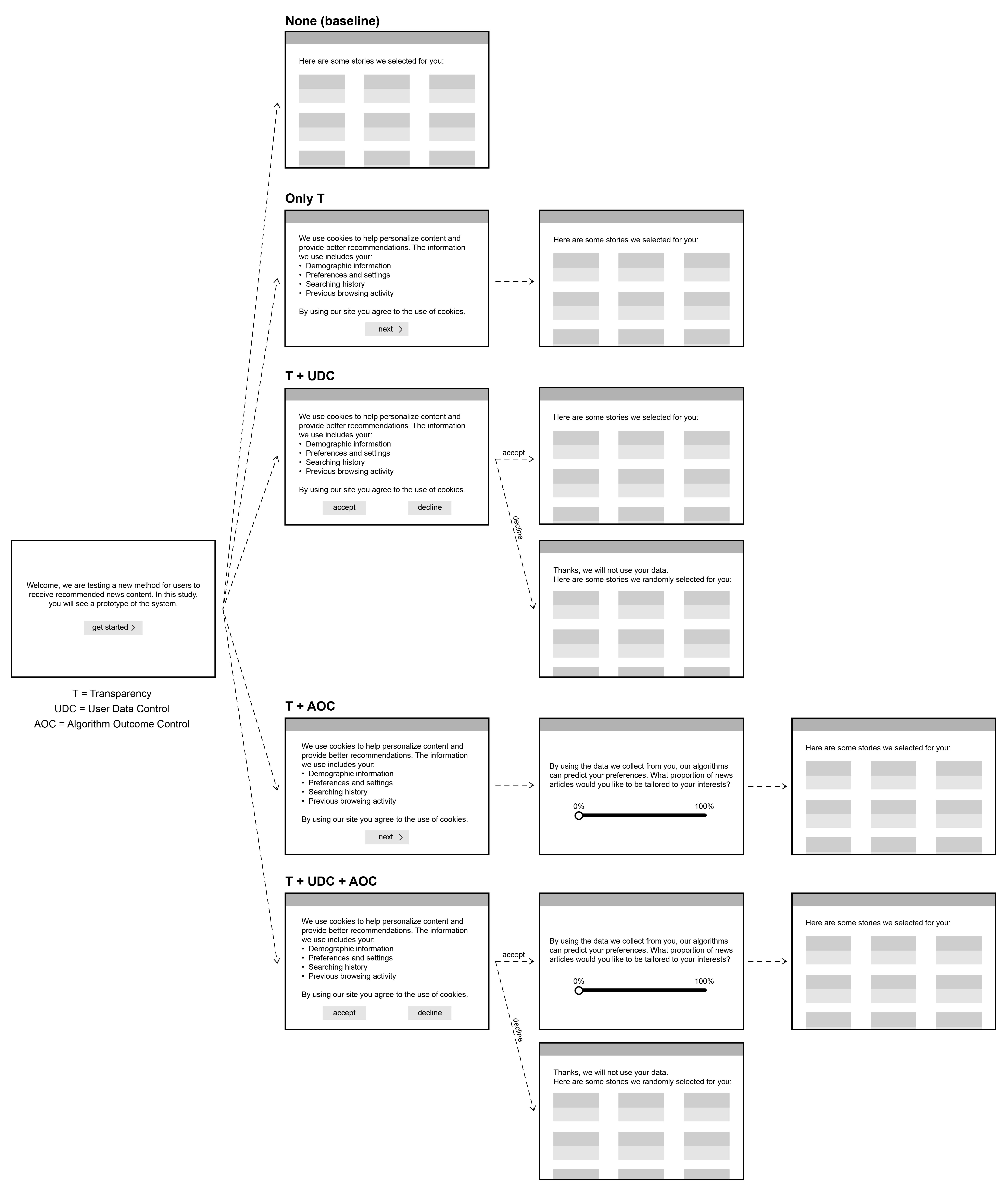}
    \caption{Wireframe of the Recommendation Flows}
    \label{fig:wireframe}
\end{figure*}

\begin{figure*}
    \centering
    \includegraphics[width=\linewidth]{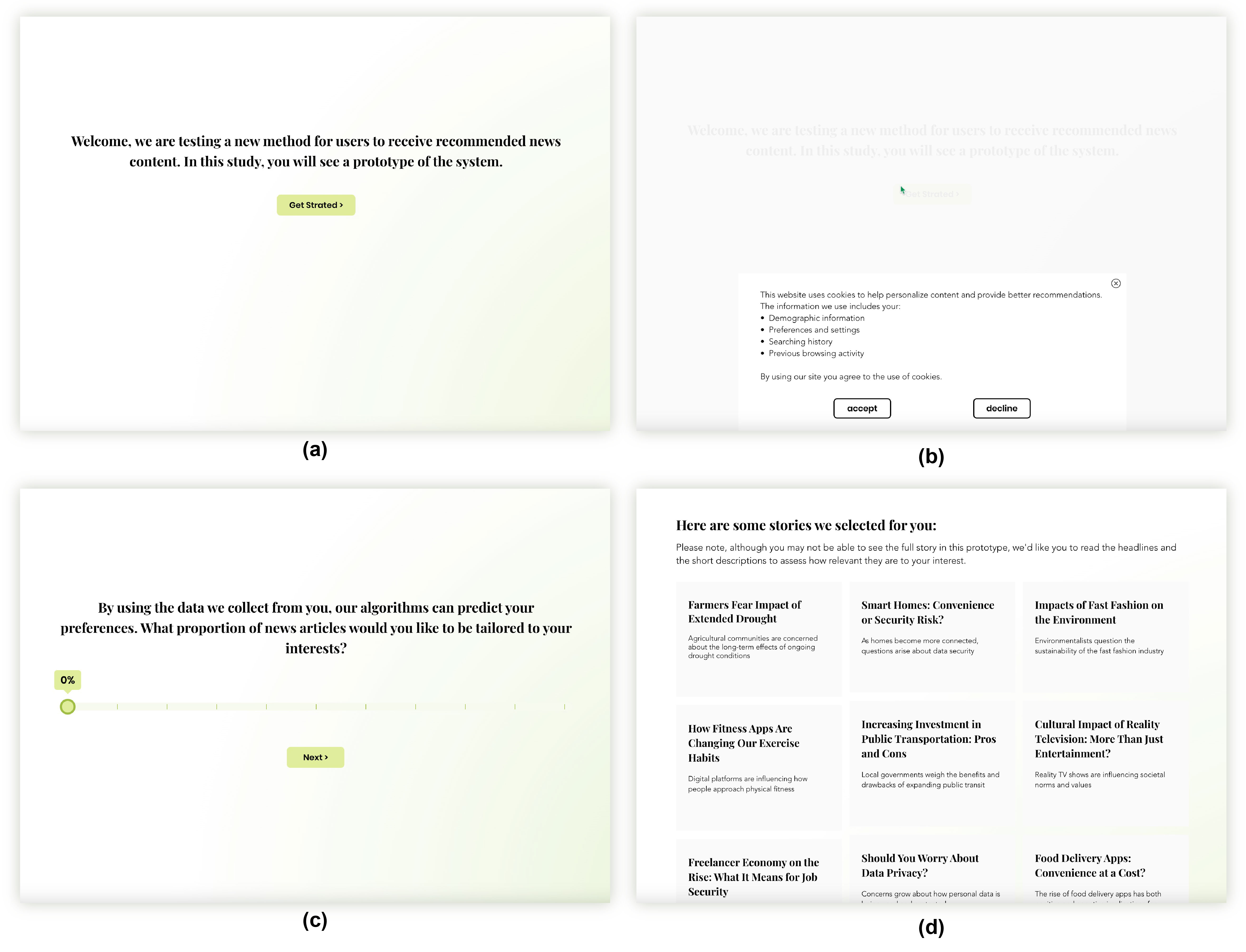}
    \caption{Representative Finalized Webpages: (a) Landing Page, (b) User Data Control (UDC), (c) Algorithm Outcome Control (AOC), (d) Recommendation Result Page.}
    \label{fig:sample}
\end{figure*}

\end{document}